\documentclass[reprint,a4paper,nofootinbib,superscriptaddress,showpacs,floatfix]{revtex4-1}
\usepackage{fullpage}
\usepackage[english]{babel}
\usepackage[OT2,T1]{fontenc}

\usepackage{mathrsfs}
\usepackage{amsmath}
\usepackage{amssymb}
\usepackage{epsfig}
\usepackage{marvosym}
\usepackage{subfigure}
\usepackage{fancyhdr}
\usepackage{rotating}
\usepackage{graphicx}
\usepackage[pdftex]{hyperref}
\hypersetup{colorlinks=false}
\usepackage{tikz}
\usepackage{verbatim}
\begin{document}
\author{J. Sweep}
\affiliation{Radboud University Nijmegen, Institute for Molecules and Materials, Heyendaalseweg 135, NL-6525AJ Nijmegen, The Netherlands}
\author{A. N. Rubtsov}
\affiliation{Department of Physics, Moscow State University, 119992 Moscow, Russia}
\author{M. I. Katsnelson}
\affiliation{Radboud University Nijmegen, Institute for Molecules and Materials, Heyendaalseweg 135, NL-6525AJ Nijmegen, The Netherlands}
\date{\today}
\title{Non-local correlation effects and metal-insulator transition in the s-d exchange model}
\begin{abstract}
The metal-insulator transition within the s-d exchange model is studied by the Dynamical Mean Field Theory and Dual Fermion approaches. The latter takes into account nonlocal correlation effects which are shown to be essential. In particular, the critical values of the s-d exchange coupling constant for these two methods turn out to be different by a factor of more than two (for the case of square lattice and spin 1/2 localized electrons). The calculations were performed using a Continuous-Time Quantum Monte Carlo method. The difference of the quantum spin-1/2 case and the classical-spin case is discussed. For the quantum case the sign of s-d exchange coupling constant is relevant which demonstrates an importance of the Kondo effect.
\end{abstract}
\maketitle
\section{Introduction}
The motivation to study the metal-insulator transition in the s-d exchange, or Vonsovsky-Zener, model \cite{vons,zener} (now it is frequently called Kondo lattice model) is given by previous successes obtained studying the metal-insulator transition (MIT) in the Hubbard model \cite{HubbardI}. The s-d exchange model is complementary to the Hubbard model. While in the Hubbard model the interacting electrons responsible for magnetism and for electron conductivity are located in the same band, in the s-d exchange model they are located in different bands. In one band the electrons are delocalized or itinerant, therefore these electrons are called conduction electrons. The other band contains electrons that are localized on lattice sites. The type of the interaction is also different between the two models. The Hubbard model contains an on-site Coulomb interaction modeled by a single energy parameter $U$. The s-d exchange model contains a spin-spin interaction between the localized and conduction electrons with a coupling constant (s-d exchange parameter) $J$. The s-d exchange model is used in the description of rare earths, in which 4f-electrons have a very localized nature \cite{Magnetism}. The description of colossal magnetoresistance materials such as (La,Ca,Sr)MnO compounds sometimes is based on the s-d exchange model assuming that the $t_{2g}$ electrons are rather localized and the $e_g$ electrons are itinerant \cite{Nagaev}.

Whereas for relatively weakly correlated systems the crystal with half-filled energy bands is a metal, for strong enough correlation the system should be an insulator, with each electron sitting in a localized state associated with a given site. This corresponds to the splitting of the energy bands into Hubbard subbands separated (for a large enough interaction parameter $U$) by a gap \cite{Mott,Imada}. In the s-d exchange model the parameter $U$ is replaced by $J\cdot S$ where $S$ is the spin of localized electrons. The Hubbard-III approximation \cite{Hubbard,Ano91} provides a qualitative description of the metal-insulator transition for both models. This description however does not take into account all quantum effects; for the case of s-d exchange model, it turns out to correspond to the classical spin limit $S \rightarrow \infty$ \cite{Ano91}. The Dynamical Mean Field Theory (\textsc{dmft}) \cite{Georges} was a real breakthrough in this respect dramatically improving the description of metallic phase near the transition. Whereas within the Hubbard-III approximation the main effect of approaching the MIT is the growth of damping of electron states at the Fermi energy, together with appearance of shoulders as the precursors of future Hubbard bands (two-peak structure), the \textsc{dmft} gives the {\it three}-peak structure, with the ``Kondo peak'' near the Fermi energy, that is, the growth of effective mass instead of the growth of the damping. The origin of this peak is the formation of a resonance due to electron scattering by {\it quantum} spins \cite{Georges}. The dynamical mean-field theory is formally exact in the limit of infinite space dimensionality whereas the Hubbard-III approximation is formally exact (within the s-d exchange model) in the limit of infinite space dimensionality {\it and} infinite, that is, classical, spin \cite{Ano91}. In particular, in this limit there is no difference between the case of ferromagnetic ($J<0$) and antiferromagnetic ($J>0$) s-d exchange coupling; the Kondo resonance exists only in the latter case (the detailed discussion of the Kondo effect and related physics can be found in Ref.\onlinecite{Hewson}).

The \textsc{dmft} is the best local approximation, that is, it corresponds to an optimal self-consistent choice of the momentum-independent electron self-energy. For real two- and three-dimensional systems it is just an approximation, which is not quite controllable.  The Dual Fermion formalism \cite{Brief_report,Dual Fermion09,Dual_boson} allows to construct a diagrammatic expansion for the non-local correlation effects starting from the \textsc{dmft} Green's function as a zeroth-order approximation. The differences between the Dual Fermion results and the \textsc{dmft} results were investigated recently \cite{Andrey} for the Hubbard model on triangular lattice, and turned out to be very essential. The present paper discusses the differences between Dual Fermions and \textsc{dmft} for the metal-insulator transition in the s-d exchange model.

The MIT in the s-d exchange model in the classical limit $S \rightarrow \infty$ has been studied in Ref.\onlinecite{Ano91}. In this work we consider a \textit{quantum} s-d exchange model at a two-dimensional (2D) square lattice, for $S=\frac{1}{2}$ where we expect quantum effects to be maximal. We study the role of non-local correlation effects by comparison of the results of \textsc{dmft} and Dual Fermion calculations. We show that the \textsc{dmft} strongly overestimates the critical value of the coupling constant for the metal-insulator transition $J_c$. We also show that in the s-d exchange model quantum effects are indeed relevant, since the  results for positive and negative values of the coupling constant $J$ are essentially different.

\section{Hamiltonian}
The s-d exchange model contains electrons that are localized on lattice sites, and conduction electrons that can hop between different sites $i,j$. Let us introduce their annihilation operators $c_{i\sigma}$, $\sigma=\uparrow,\downarrow$ is the spin projection. The annihilation operators for the localized electrons will be designated $d_{i\sigma}$. The Hamiltonian contains (in order of appearance in eq. \eqref{H}) a hopping term, a Hubbard term that is included for calculational purposes, the s-d exchange interaction, and the chemical potential $\mu$ term:
\begin{widetext}
\begin{equation}
H=-\sum_{ij\sigma} t_{ij}(c^{\dag}_{j\sigma}c_{i\sigma}) - \sum_i U(n_{li\uparrow}-n_{li\downarrow})^2 + J\sum_{i\sigma} \vec{S_i} \cdot \vec{s_i} + \mu N,
\label{H}
\end{equation}
$N$ is the number of conduction electrons and $n_{li\sigma}=\sum_{\sigma}d^{\dag}_{i\sigma}d_{i\sigma}$.
The spin-spin interaction term reads:
\begin{align}
J& \vec{S_i} \cdot \vec{s_i}=\frac{J}{4}(n_{l\uparrow}-n_{l\downarrow})(n_{c\uparrow}-n_{c\downarrow})+\frac{J}{2}d^{\dag}_{\uparrow}d_{\downarrow}c^{\dag}_{\downarrow}c_{\uparrow}+\frac{J}{2}d^{\dag}_{\downarrow}d_{\uparrow}c^{\dag}_{\uparrow}c_{\downarrow}.
\end{align}
\end{widetext}
The Hubbard interaction is included to (computationally) ensure the half-filling of the lattice sites (one localized electron per site). Strictly speaking, it is reached only in the limit $U\to\infty$. We will do the calculations for a large but finite $U$. With this Hubbard-term and $U\gg J$ and $\beta U\gg 1$ half-filling is achieved to sufficient extent.

\section{Method}
Our goal was to study the density of conduction-electron states (DoS) of the s-d exchange model at different coupling constants $J$. The calculations were carried out at the Matsubara axis (imaginary frequencies) and, to obtain the densities of states we performed an analytical continuation of the Green's functions on the real axis, using the Maximum Entropy method \cite{Maxent, Sandvik}. The Green's functions we obtained self-consistently in \textsc{dmft} versus the Dual Fermion approach.
In short, both \textsc{dmft} and Dual Fermion utilize an effective-medium concept, so that they deal with the impurity model described by the action $S_{imp}=S_{at}+\Delta_\omega c^\dag_{j, \omega} c_{\omega,j}$, where $S_{at}$ is the action of an isolated lattice site, and $\Delta$ is a hybridization by the effective bath. Frequency-dependence of the hybridization allows to handle the effects of local fluctuations, that underlie the Kondo physics. On the other hand,  this $\omega$-dependence makes the impurity problem non-hamiltonian and therefore requires a usage of numerical solvers to obtain its Green function $g_\omega$ and two-particle vertex $\gamma^{(4)}$ (the later is required for Dual Fermions).  As a numerical solver, we used the interaction-expansion continuous-time quantum Monte Carlo procedure \cite{ARcode,RMPP}. The Dual fermion lattice Green's function equals (spin and orbital indices are omitted)
\begin{equation}
G_{\omega, k}=\frac{1}{(g_\omega+g_\omega \tilde \Sigma_{\omega k} g_\omega)^{-1}+\Delta_\omega-\epsilon_k},
\end{equation}
where $\epsilon_k$ is a dispersion law arising from the hopping $t_{ik}$, and $\tilde \Sigma$ is the self-energy function of the dual ensemble obtained in certain diagrammatic approximation. Vertices of dual diagrams are vertex parts of the impurity problem.
\textsc{dmft} can be represented as the zeroth-order Dual Fermion scheme with  $\tilde \Sigma=0$. In this approximation, 
the lines (propagators) in dual diagrams have a simple meaning, as they are equal to the non-local part of the Green's function, $\tilde G_{\omega,k}^{DMFT}= G_{\omega,k}- G_{\omega,r=0}$ (for a general case, $\tilde G$ and $G$ are also exactly related, although the relationship is more complicated). One can easily see that \textsc{dmft} describes purely local contribution to the lattice self-energy (and is actually shown to correspond to the summation of all local diagrams), whereas Dual Fermion diagrams describe non-local effects. More technical detail can be found in Refs. \cite{Brief_report,Dual Fermion09}.

The hybridization function $\Delta_\omega$ was obtained self-consistently from the condition that the  local part of the dual propagator vanishes, $\tilde G_{\omega, r=0}=0$. This eliminates the first-order Hartree dual diagram, and thus the leading correction to $\tilde \Sigma$ is given by the second-order (see diagram (b) and formula (20) of Ref.\cite{Dual Fermion09}),
\begin{equation}
\tilde\Sigma_{\omega, r}=\frac{1}{2 \beta^{2}} \sum_{\omega+\omega_2=\omega'+\omega_2'} \gamma^{(4)}_{\omega \omega' \omega_2 \omega_2'} \tilde G_{\omega' r} \tilde G_{\omega_2 r} \tilde G_{\omega_2' r}\gamma^{(4)}_{\omega' \omega \omega_2' \omega_2} 
\end{equation}
(summation over spins is also implied).

The last expression is an approximation used for the Dual Fermion calculations in the present paper. The calculations were performed for the paramagnetic state of the Hamiltonian (\ref{H}) on a square lattice with the nearest-neighbor hopping parameter $t=0.25$ (which means a half-bandwidth equal to 1), we have chosen $U=5$ and the inverse temperature $\beta=10$. Dual fermion diagrams were calculated at a $24\times 24$ slab in momentum space.
The final accurate calculations took about two weeks of CPU load per $J$-value in the \textsc{dmft} case and about 10\% more in the Dual Fermion case. The amount of self-consistency loops was roughly 10-20 while the number of Monte Carlo steps was roughly doubled with each loop in the last 6 loops until it reached $8 \cdot 10^8$.

\section{Results}
\subsection{Critical coupling constant for metal-insulator transition}
\begin{figure*}[!!h!!]
\centering
\includegraphics[width=0.9\textwidth]{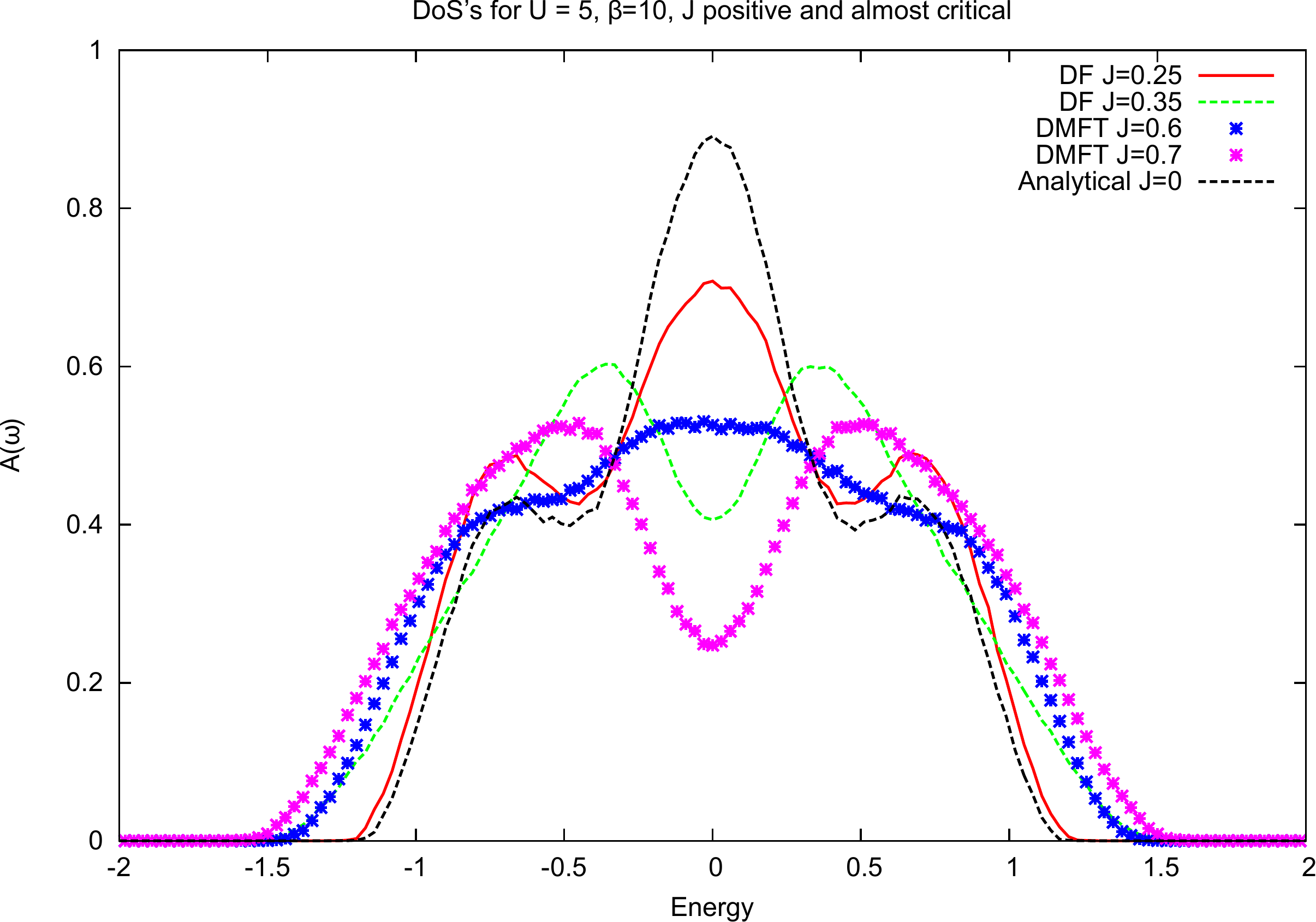}
\caption{The DoS around the critical values of $J$. The stars correspond to \textsc{dmft} results, the lines are Dual Fermion calculations. We see the metal-insulator taking place as the local maximum (peak) at the Fermi level ($E=0$) disappears for higher $J$-values. The values are chosen such that they are just around the transition for the two calculational schemes, \textsc{dmft} and Dual Fermions. For \textsc{dmft} the value $J_c\approx0.65$, and for the Dual Fermion calculation it gets as low as $J_c\approx0.3$.}
\label{Jcrit}
\end{figure*}

The MIT for the s-d exchange model is visible in Fig. \ref{Jcrit}. We have quoted values of the coupling constant just around the transition. Although the temperature is quite high and the Hubbard-$U=5$ we still see a qualitatively correct picture of the densities of states, accurate enough to estimate a reasonable value of the coupling constant $J$ for the metal-insulator transition. We consider the state to be conducting if there is a local maximum at the Fermi level and insulating if there is a local minimum at the Fermi level, because the temperature is so high.

The choice of $U=5$ in the Quantum Monte Carlo calculations is justified because in our calculations $\beta U=50\gg1$ and $U/J\geq5$ so that half-filling is achieved to sufficient extent.

\begin{table*}
\begin{tabular}{l|l|l}
Method & $J_c$ &Features \\
\hline
\hline
Simple self-consistent \cite{Ano91}  & $0.816$& Solution of Heisenberg equation of motion, classical limit: $S\to\infty$\\
Hubbard-III \cite{Ano91} & $\approx0.8$& Inclusion of scattering corrections\\
\textsc{dmft} [this work]& $\approx0.65$& Quantum calculation, local correlations\\
Dual Fermions [this work]&$\approx0.3$& Quantum calculation, fermionic non-local corrections\\
\end{tabular}
\caption{Critical values of $J$ for the metal-insulator transition in different approaches.}
\label{Jcrit_table}
\end{table*}

In Table \ref{Jcrit_table} we compare several different calculational methods that were used to obtain a critical value of the coupling constant for the metal-insulator transition.

\subsection{Difference for positive and negative coupling constant}
\begin{figure*}[h!!!]
\includegraphics[width=0.9\textwidth]{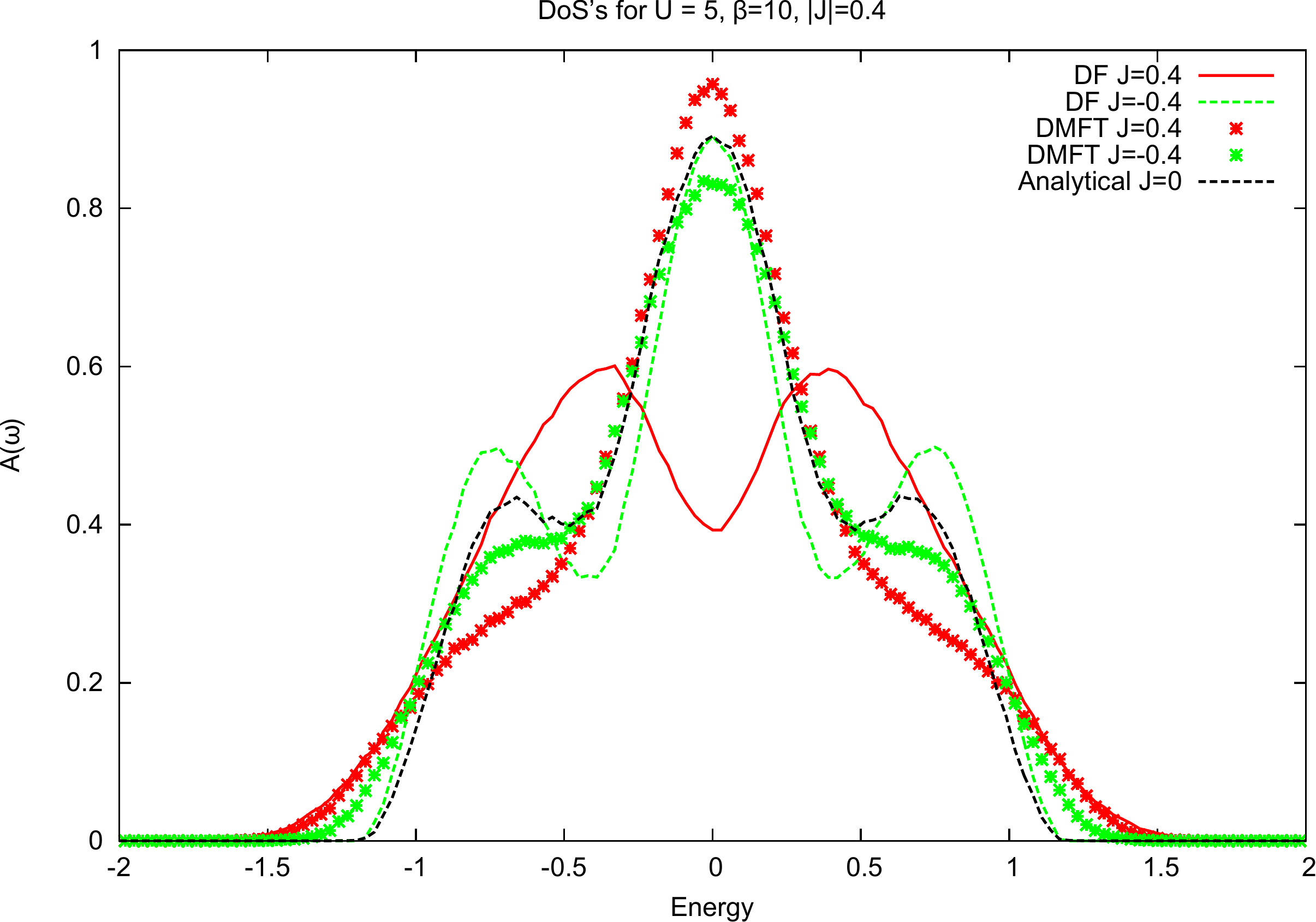}
\caption{DoS's for $|J|=0.4$. There is a difference for positive and negative $J$. For the local theory, \textsc{dmft} (stars), the difference is already visible. Beyond this approximation, incorporating non-local correlations using the Dual Fermion approach (lines), it is more pronounced.}
\label{Jsmall}
\end{figure*}

\begin{figure*}[h!!]
\includegraphics[width=0.9\textwidth]{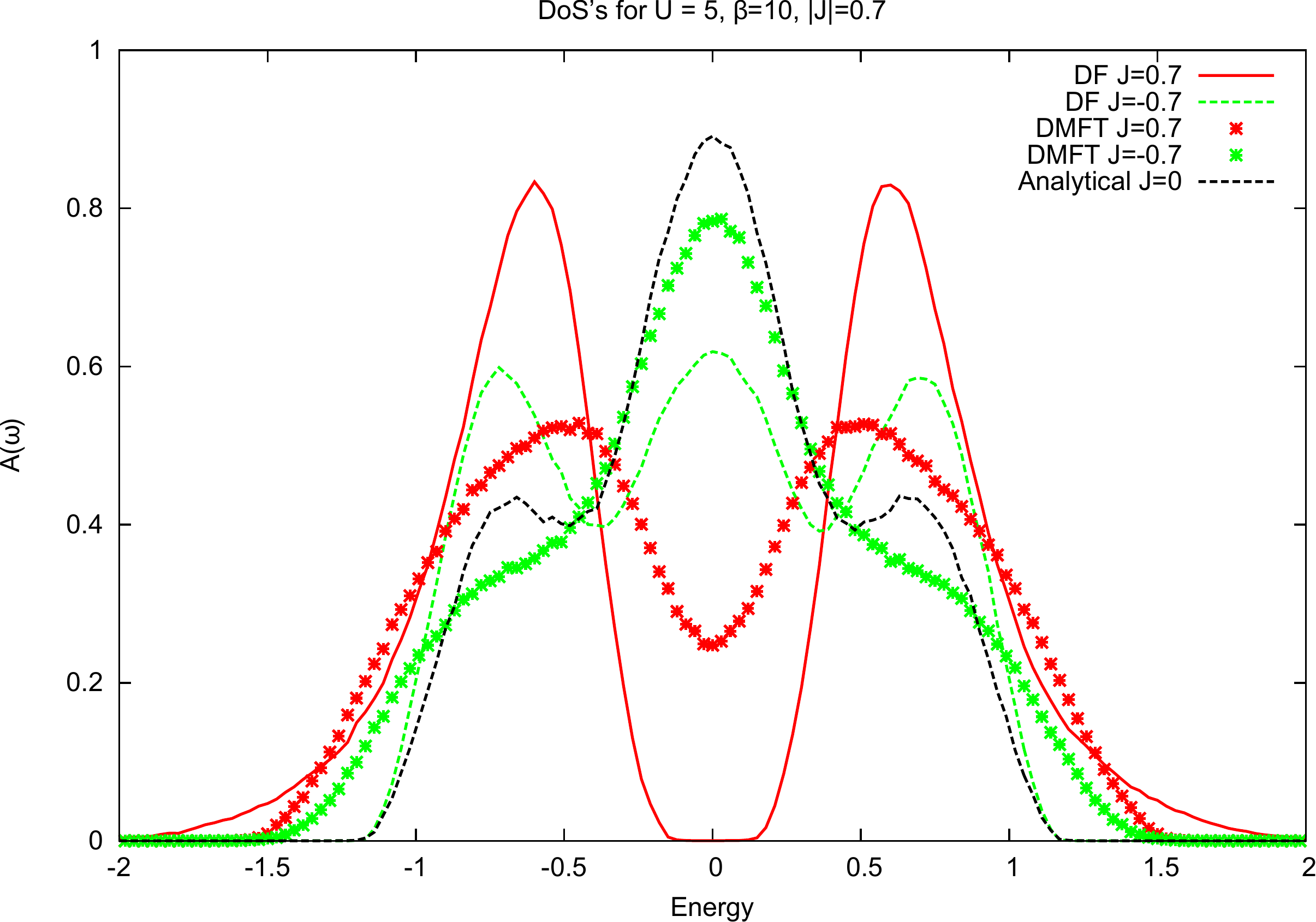}
\caption{DoS's for $|J|=0.7$. In this regime the difference between positive and negative $J$ is clearly visible in both the local (\textsc{dmft}, stars) and the non-local (Dual Fermion, lines) approaches.}
\label{Jlarge}
\end{figure*}

The difference in behavior of the DoS between positive and negative values of the coupling constant is displayed in Figures \ref{Jsmall} and \ref{Jlarge}

In Figure \ref{Jsmall} we display the results for $|J|=0.4$ calculated within both \textsc{dmft} and Dual Fermions. The different colors represent positive and negative $J$-values. Stars correspond as before to the \textsc{dmft}-calculation while the Dual Fermion result is drawn with lines. In this picture we look at the general shape of the curves. We see that they are not the same for positive and negative $J$ and that incorporating non-local corrections through Dual Fermions changes the results: the difference is more pronounced in the Dual Fermion case.

In Figure \ref{Jlarge} we display the results for $|J|=0.7$, also calculated using \textsc{dmft} (stars) and Dual Fermions (lines). The different colors again represent the positive and negative $J$-values. The general shape of the curves is much more different between positive and negative $J$ for $|J|=0.7$ than for $|J|=0.4$.

\section{Conclusions}
In this work we justify two claims. First, the Dual Fermion formalism description of the metal-insulator transition in the s-d exchange model to date predicts a critical value of the coupling constant $J_c\approx0.3$. This value differs from the \textsc{dmft}-result by a factor of more than 2 and corresponds nicely to the result obtained in the other work using a non-local correlations approach \cite{KondoAux01}. The different results between various calculational schemes are summarized in table \ref{Jcrit_table}.

Second, there is a clear qualitative difference between densities of states of the s-d exchange model for antiferromagnetic and ferromagnetic s-d exchange coupling. This shows that incorporating quantum effects such as the Kondo effect is essential for a good description of the model.

The results in this work were part of the Master's thesis work of JS.

\begin{acknowledgements}
The Dual Fermion code was used with kind permission from H. Hafermann. The authors thank A. Lichtenstein for hospitality and fruitful discussions during their stay in Hamburg.
\end{acknowledgements}

\end{document}